\documentclass[12pt,a4paper,reqno]{article}
\usepackage[left=2cm,right=2cm,top=2cm,bottom=2cm]{geometry}

\usepackage[utf8]{inputenc}
\usepackage[english]{babel}
\usepackage{amsmath}
\usepackage{amsfonts}
\usepackage{amssymb}
\usepackage[titletoc,title]{appendix}
\usepackage{mathtools}
\usepackage{bbm}
\usepackage{bm}
\usepackage{subdepth}
\usepackage{enumerate}
\usepackage{mathrsfs}
\usepackage{graphicx}
\usepackage[outdir=./]{epstopdf}
\usepackage{esint}
\usepackage{dsfont}
\usepackage{amsthm}
\usepackage{cite}
\usepackage{url}
\usepackage{hyperref}
\usepackage{cleveref}
\hypersetup{%
        colorlinks=true,
        citecolor={black},
        linkcolor={black},
        urlcolor={black},}
\newcommand{\nm}{\noalign{\smallskip}}

\newcommand*{\RR}{\mathbb{R}}

\newcommand*{\smin}{\setminus}

\newcommand*{\e}{\varepsilon}

\newcommand*{\pd}{\partial}

\newcommand*{\vphi}{\varphi}

\newcommand{\G}{\Gamma}

\theoremstyle{plain}
\newtheorem{thm}{Theorem}[section]
\newtheorem{proposition}[thm]{Proposition}
\newtheorem{lemma}[thm]{Lemma}

\newtheorem{definition}[thm]{Definition}
\newtheorem{remark}[thm]{Remark}

\theoremstyle{definition}

\theoremstyle{remark}

\renewcommand{\o}{\omega}
\renewcommand{\l}{\lambda}
\renewcommand{\O}{\Omega}

\newcommand*{\bra}[1]{\left\lbrace #1 \right\rbrace}

\newcommand{\cqfd}{\hfill $\blacksquare$\\ \medskip}
\makeatletter
\def\blfootnote{\gdef\@thefnmark{}\@footnotetext}
\makeatother

\begin{document}
\title{Perturbations of the scattering resonances of an open cavity by small particles. Part II: The transverse electric polarization case}

\author{
Habib Ammari\thanks{\footnotesize Department of Mathematics, 
ETH Z\"urich, 
R\"amistrasse 101, CH-8092 Z\"urich, Switzerland (habib.ammari@math.ethz.ch, alexander.dabrowski@sam.math.ethz.ch).} \and 
Alexander Dabrowski\footnotemark[1] \and Brian Fitzpatrick\thanks{\footnotesize  
ESAT - STADIUS,
Stadius Centre for Dynamical Systems,
Signal Processing and Data Analytics,
Kasteelpark Arenberg 10 - box 2446,
3001 Leuven,
Belgium (bfitzpat@esat.kuleuven.be).}  \and Pierre Millien\thanks{\footnotesize  Institut Langevin,  1 Rue Jussieu, 75005 Paris, France (pierre.millien@espci.fr).} }

\date{} 

\maketitle

\begin{abstract}
This paper is concerned with the scattering resonances of open cavities. It is a follow-up of \cite{paper1}, where the transverse magnetic polarization was assumed. In that case,
 using the method of matched asymptotic expansions, the leading-order term in the shifts of scattering resonances due to the presence
of small particles of arbitrary shapes was derived and the effect of radiation on the perturbations of open cavity modes was characterized. The derivations were formal. 
 In this paper, we consider the transverse electric polarization and prove a small-volume formula for the shifts in the  scattering resonances of a  radiating dielectric cavity perturbed by small particles. We show a strong  enhancement in the frequency shift in the case of plasmonic particles. We also consider exceptional scattering resonances and perform small-volume asymptotic analysis  near them.  Our method in this paper relies on pole-pencil decompositions of  volume integral operators.  
\end{abstract}

\def\keywords2{\vspace{.5em}{\textbf{Mathematics Subject Classification
(MSC2000).}~\,\relax}}
\def\endkeywords2{\par}
\keywords2{35R30, 35C20.}

\def\keywords{\vspace{.5em}{\textbf{Keywords.}~\,\relax}}
\def\endkeywords{\par}
\keywords{Open dielectric resonator, shift of scattering resonances, plasmonic nanoparticles, exceptional scattering resonances, pole-pencil decomposition.}

\section{Introduction}

In this paper, which is a follow-up of \cite{paper1}, we consider dielectric radiating cavities \cite{asp,cao,maier} and rigorously obtain asymptotic formulas 
for the shifts in the scattering resonances that are due to a small particle of arbitrary shape. Our formula shows that the perturbations of the scattering resonances can be expressed in terms of the polarization tensor of the small particle. The scattering resonances  can be degenerate or even exceptional and the small particle can be plasmonic. Our method is based on pole-pencil decompositions (see, for instance, \cite{ammari2017mathematical, surv}) of the volume integral operator associated with the radiating dielectric cavity problem. The new technique introduced in this paper can not be easily extended to the transverse magnetic case considered in \cite{paper1} due to the hyper-singular character of the associated volume-integral operator.  

The paper is organized as follows. In Section \ref{sec-1p2}, we characterize the scattering resonances of dielectric cavities in terms of the spectrum of a volume integral operator. In Section \ref{sec-2p2},  
using the method of pole-pencil decompositions, we derive the leading-order term in the shifts of scattering resonances of an open dielectric cavity due to internal particles. 
In Section \ref{sec-3p2}, using a Lippmann-Schwinger representation formula for the Green's function associated with the open cavity, we generalize the formula obtained in Section \ref{sec-2p2} to the case of external particles. In Section \ref{sec-4p2}, we consider the perturbation of an open dielectric cavity by plasmonic nanoparticles.  The formula obtained for the shifting of the frequencies shows a strong  enhancement in the frequency shift in the case of plasmonic nanoparticles.  In Section \ref{sec-5p2}, we 
perform an asymptotic analysis for the shift of exceptional scattering resonances. The paper ends with some concluding remarks.

\section{Scattering resonances of a dielectric cavity} \label{sec-1p2}

\subsection{Model}
We consider the scattering of linearly polarized light by a dielectric cavity in a time-harmonic regime. Let $\Omega$ be a bounded domain in $\mathbb{R}^d$ for $d=2,3,$ with smooth boundary $\partial \Omega$.  Assume $\e \equiv \tau \e_c + \e_m$ inside $\O$ and $\e = \e_m$ outside $\O$, and $\mu = \mu_m$ everywhere. Here, $\e_c,\e_m,$ and $\tau$ are positive constants. Since we are interested in scattering resonances, we look for solutions $u$ of the homogeneous Helmholtz equation at frequency $\omega$:

\begin{align}\label{eq:helmholtz}
\left\{\begin{aligned}
&\Delta u  + \omega^2 \varepsilon(x) \mu_m  u =0  \qquad \text{in} \  \mathbb{R}^d ,\\
&u \text{\ satisfies the outgoing radiation condition.} 
\end{aligned}\right.
\end{align}

Let $\G_m$ be the outgoing fundamental solution of $\Delta + \e_m \mu_m \o^2$ in  free space, and let $G$ be the outgoing fundamental solution of $\Delta + \e \mu_m \o^2$ in free space.
We define the following integral operator:

\begin{definition} Let
\begin{align*}
L^2(\Omega) \longrightarrow & L^2(\Omega) \\
u \longmapsto &  K_\O^\o [u]:=- \int_\O u(y) \G_m (\, \cdot \, -y;\omega) d y.
\end{align*}
\end{definition}

The following Lippmann-Swchinger representation formula holds: 

\begin{proposition}
$u$ is a solution of \eqref{eq:helmholtz} if and only if $u$ is a solution of
\begin{align}\label{eq:LSE}
\left( I - \omega^2 \tau \varepsilon_c \mu_m  K_\O^\o \right) [u] = 0.
\end{align}
\end{proposition}
According to \cite{hai}, the following spectral decomposition of the operator $K_\O^\o$ holds:
\begin{lemma} \label{lem23} The operator $K_\O^\o$ is bounded from $L^2(\Omega)$  into $H^2(\Omega)$. Moreover, it is a Hilbert-Schmidt operator. Therefore, its spectrum is $$\sigma(K_\O^\o) = \bra{0, \l_1(\o), \l_2(\o), \ldots, \l_j(\o), \ldots },$$ where $|\l_j(\o)| \rightarrow 0$ as $j\rightarrow +\infty$ and $\{0\} = \sigma(K_\O^\o) \setminus \sigma_p(K_\O^\o)$ with $\sigma_p(K_\O^\o)$ being the point spectrum.
\end{lemma}

\begin{remark} \label{rem24} The scattering resonances are precisely the frequencies for which $\left(\omega^2 \tau \varepsilon_c \mu_m\right)^{-1}$ belongs to the spectrum of $K_\Omega^\omega$.
\end{remark}

\begin{remark} Note that  $\Im \lambda_j(\omega) \neq 0$ for all $j$ and $\omega \in \mathbb{R}$.  
\end{remark}

Let $H_j$ be the generalized eigenspace associated with $\l_j(\o)$.
Then, from \cite{hai},  it follows that $L^2(\O) $ is the closure of $\bigcup_j H_j$.
 \begin{lemma} We have
$$L^2(\Omega) = \overline{\bigcup_j H_j}.$$
\end{lemma}

\begin{lemma}\label{lem:completeness} Assume that for any $j$, $\text{dim\ }H_j=1$,  and denote by $e_j$ a unitary basis vector for $H_j$. Then the functions \begin{align*}
f_{j,k}(x,y) = e_j(x)e_k(y),
\end{align*} form a normal basis for $L^2(\Omega \times \Omega)$. Moreover, 
\begin{align*}
\delta(x-y) = \sum_j e_j(x) e_j(y).
\end{align*}
\end{lemma}
\subsection{Pole pencil decomposition of the Green's function}

We denote by $ G(x,y; \omega) $ the Green's function associated with problem \eqref{eq:helmholtz}, that is, the solution in the sense of distributions of
\begin{align*}
\left(\Delta_x  + \omega^2 \varepsilon(x) \mu_m \right)G(x,y,\omega) = \delta_y  
\end{align*}
satisfying the outgoing radiation condition.

\begin{definition}\label{de:nonexceptionnal} In view of Lemma \ref{lem23} and Remark \ref{rem24}, we say that $\omega_0$ is a scattering resonance for the open cavity problem if there exists a $j_0$ such that 
\begin{equation}
\label{eqp2-1} 
1  - \o_0^2 \tau \e_c \mu_m \l_{j_0} (\o_0) = 0.\end{equation} 
We say that the scattering resonance $\o_0$ is a non-exceptional scattering resonance if the following assumptions hold:
\begin{itemize}
\item[(i)]  We have $$1 - \o^2 \tau \e_c \mu_m \l_{j_0} (\o) = R(\omega) (\o - \o_0),$$
where $R(\omega_0) \neq 0$ and $\omega \mapsto R(\omega)$ is analytic;
\item[(ii)]  The generalized eigenspace $H_{j_0}(\omega)$ is of dimension $1$.
\end{itemize}
\end{definition}

\begin{remark}It is easy to see that for $\tau$ large enough, (\ref{eqp2-1}) has solutions.  
\end{remark}

We can now give the following expansion for $G$ when $\omega$ is close to a non exceptional scattering resonance. We refer to Appendix \ref{appendix0} for its proof. 

\begin{proposition}\label{prop:polepencilG}
Assume that $\o_0$ is a non-exceptional scattering resonance.
There exists a complex neighborhood $V(\omega_0)$ of $\omega_0$ such that for $\omega$ in $V(\omega_0) \setminus \{\omega_0\}$, 
\begin{equation}\label{eq:polepencil}
G(x,y; \omega) =   \G_m(x-y;\omega) + c_{j_0}(\omega_0) \dfrac{e_{j_0}(x;\omega) e_{j_0} (y;\omega)}{\o - \o_0} + \tilde{R}(x,y,\omega), 
\end{equation}
where $\mathrm{vect}(e_{j_0})= H_{j_0}$. Moreover,  $\omega \mapsto\tilde{R}(x,y,\omega)$, $\omega \mapsto e_{j_0}(\,\cdot, \omega)$, and 
 $\omega \mapsto c_{j_0}(\omega)$ are all analytic  in $V(\omega_0)$, and $(x,y) \mapsto \tilde{R}(x,y,\omega)$ is smooth.
\end{proposition}

\section{Shift of the scattering resonances by internal small particles} \label{sec-2p2}

Now let $D \Subset \O$ be a small particle of the form $D= z+\delta B$, where $\delta$ is the characteristic size of $D$, $z$ is its location, and $B$ is a smooth bounded domain containing the origin.  We suppose that $D$ has a  magnetic permeability that is different from $\mu_m$, and consider the operator
$$\nabla \cdot \dfrac{1}{\mu} \nabla + \e \o^2, $$
where $\mu = \mu_c$ in $D$ and $\mu= \mu_m$ outside $D$.

As $\delta \rightarrow 0$, we seek an $\o_\delta$ in a neighborhood of $\o_0$ such that there exists a non-trivial solution to
\begin{equation} \label{eq:perturbed} (\nabla \cdot \dfrac{1}{\mu} \nabla + \e \o^2_\delta ) u = 0,\end{equation} subject to the outgoing radiation condition.

The following asymptotic expansion of $\omega_\delta$ holds.
\begin{proposition} \label{prop:perturb1}
As $\delta \rightarrow 0$, we have
\begin{equation} \label{paper2-form1}  \o_\delta - \o_0 \simeq \delta^d c_{j_0}(\omega_0)   M(\mu_m / \mu_c,B)  \nabla e_{j_0}(z;\omega_0) \cdot \nabla {e}_{j_0}(z; \omega_0). \end{equation}
\end{proposition}

Before proving the above result, we state the following useful lemma. We refer to Appendix \ref{appendix1} for its proof. 
\begin{lemma} \label{lemap} Let 
$$T_D^\omega : v  \mapsto \nabla_x \int_D v(y) \cdot \nabla G(x-y; \omega) dy . $$
Then, $T_D$ is a well defined operator from $L^2(D,\mathbb{R}^d)$ into itself.
\end{lemma}

\proof (of Proposition \ref{prop:perturb1})

The outgoing solution to problem (\ref{eq:perturbed}) admits the following Lippmann-Schwinger representation formula:
$$u(x) = (\frac{\mu_m}{\mu_c} - 1) \int_D \nabla u(y) \cdot \nabla G(x,y;\omega_\delta) dy \quad \mbox{for all } x \in \mathbb{R}^d.$$
Let 
$$T_D^\omega : v \in L^2(D)^d \mapsto \nabla_x \int_D v(y) \cdot \nabla G(x-y; \omega) dy \in L^2(D)^d. $$
The operator $T_D^\omega$ is well-defined, see, for instance, \cite[Appendix B]{AmmariMillien2018}.
Therefore, we seek $\omega_\delta$ such that there is a non-trivial $v \in L^2(D)^d$ satisfying
$$v(x) - (1/\mu_c - 1/\mu_m) T_D^{\omega_\delta}[v] (x) = 0 \quad \text{ for all } x \in D,$$
or equivalently, 
\begin{equation} \label{eq-p2-4}  \big(I - (\frac{\mu_m}{\mu_c} - 1)  T_D^{\omega_\delta}\big) [v]=0, \end{equation}
where $I$ denotes the identity operator.
Hence, as the characteristic size $\delta$ of $D$ goes to zero, we seek  $\omega_\delta$ in a neighborhood of $\omega_0$ such that $1/( (\mu_m/\mu_c) - 1)$ is an eigenvalue of $T_D^{\omega_\delta}$.

From the pole-pencil decomposition (\ref{eq:polepencil}) of $G$, we have
$$\nabla \int_D v \cdot \nabla G = \nabla \int_D v \cdot \nabla \G_m + \dfrac{c_{j_0} (\omega)}{\o - \o_0} \big(\int_D v \cdot \nabla {e}_{j_0}  \, dy \big) \nabla {e_{j_0}} (x;\omega) + R[v],$$
where $R: L^2(D)^d \rightarrow L^2(D)^d$ is an operator with smooth kernel that is analytic in $\o \in V(\o_0).$
Let $$N_D^\omega : v \in L^2(D)^d \mapsto \nabla_x \int_D v(y) \cdot \nabla \G_m(x-y;\omega) dy \in L^2(D)^d.$$ Then, it follows that
$$ \begin{array}{lll} \displaystyle \frac{1}{\frac{\mu_m}{\mu_c} - 1} \bigg(I - (\frac{\mu_m}{\mu_c} - 1) T_D^\omega \bigg)[v] & = & \displaystyle \bigg( \frac{I}{\frac{\mu_m}{\mu_c} - 1}  - N_D^\omega \bigg)[v] \\
\nm && \displaystyle -  \dfrac{c_{j_0}(\omega)}{\o - \o_0}  (v, \nabla {e}_{j_0}) \nabla e_{j_0} + R[v], \end{array}$$
where $(\, \cdot , \cdot \,)$ denotes the $L^2$ real scalar product on $D$.   

Let $L = 1/( (\mu_m/\mu_c) - 1) I  - N_D^{0}$, where $N_D^0:= N_D^{\omega=0}$. Then, (\ref{eq-p2-4}) can be rewritten as
$$L[v] -  \dfrac{c_{j_0}(\omega)}{\o - \o_0} (v, \nabla {e}_{j_0}) \nabla e_{j_0} + \widetilde{R}[v] = 0 ,$$
where $\widetilde{R}: L^2(D)^d \rightarrow L^2(D)^d$ is an operator with smooth kernel that is analytic in $\o \in V(\o_0).$

%From \cite{costabel2012essential, friedman1984spectral}, it is known  that $N_D^\omega|_W:  {W} \longrightarrow {W}$ is a compact operator, where $W$ is  the space of gradients of harmonic $H^1$-functions in $D$. Moreover, the spectrum of $N_D^\omega|_W$ is discrete and the associated eigenfunctions form a basis of $W$. Here, $H^1$ is the set of function in $L^2$ having their weak derivatives in $L^2$. 

Now, we need to use the orthogonal decomposition of $L^2(D, \RR^d)$ and the spectral analysis of  $N_D^0$ on $L^2(D,\RR^d)$ that can be found in \cite{costabel2012essential, friedman1984spectral}.
More precisely, recall that 
$$L^2\left(D,\RR^d\right)= \nabla H^1_0(D) \oplus {H}(\mathrm{div\ }0,D) \oplus {W},$$ where $ H^1_0(D)$ is the set of $H^1$-functions in $D$ with trace zero on $\partial D$, ${H}(\mathrm{div\ }0,D)$ is the space of divergence free $L^2$-vector fields and ${W}$ is the space of gradients of harmonic $H^1$ functions.  Here, $H^1$ is the set of function in $L^2$ having their weak derivatives in $L^2$. We will use the following lemma:
\begin{lemma}\label{lem:orthdecomp}
The operator $N_D^0$ is a bounded self-adjoint map on $L^2(D,\RR^d)$ with $  \nabla H^1_0(D)$, ${H}(\text{div\ } 0,D)$ and ${W}$ as invariant subspaces. On $\nabla H^1_0(\Omega)$, $N_D^0[\phi]=\phi$, on ${H}(\text{div\ } 0,D)$, $N_D^0[\phi]=0$ and on ${W}$: $$\nu \cdot N_D^0[\phi]= \left(\frac{1}{2} + \mathcal{K}_D^*\right)[\phi\cdot \nu] \mathrm{\ on\ } \partial D , $$
where $\nu$ is the outward normal on $\partial D$ and $K_D^*: L^2(\partial D) \rightarrow L^2(\partial D)$ is the Neumann-Poincar\'e operator associated with $\partial D$. Moreover, $N_D^0|_W:  {W} \longrightarrow {W}$ is a compact operator and hence, the spectrum of $N_D^0a|_W$ is discrete and the associated eigenfunctions form a basis of $W$.
\end{lemma}
We refer the reader to \cite{ammari2017mathematical} for the properties of the Neumann-Poincar\'e operator.  

Therefore, using Lemma \ref{lem:orthdecomp}, we have
$$v - \dfrac{c_{j_0}(\omega)}{\o - \o_0} (v, \nabla {e}_{j_0}) L^{-1}[\nabla e_{j_0}] + L^{-1} \widetilde{R} [v] = 0. $$
So, since $$|| L^{-1} \widetilde{R}||_{\mathcal{L}(L^2(D)^d, L^2(D)^d)} =o(1) \quad \mbox{ as } \delta\rightarrow 0,$$ see \cite{surv} and \cite[Lemma 4.2]{AmmariMillien2018}, the term 
$L^{-1} \widetilde{R}[v]$ can be neglected, and the following asymptotic expansion holds: 
$$\o_\delta - \o_0 \simeq c_{j_0}(\omega_0)  (L^{-1}[\nabla e_{j_0}], \nabla {e}_{j_0}).$$
Moreover,
from \cite[Proposition 3.1]{AmmariMillien2018} (see also Appendix \ref{appendix2}), it follows that 
\begin{equation} \label{eq:polarization}
(L^{-1}[\nabla e_{j_0}], \nabla {e}_{j_0}) \simeq \delta^d M(\mu_m / \mu_c,B)  \nabla e_{j_0}(z; \omega_0) \cdot \nabla {e}_{j_0}(z; \omega_0), \end{equation}
where $M$ is the polarization tensor given by \cite{kang_book}
$$
M(\mu_m / \mu_c,B) = (\frac{\mu_m}{\mu_c} -1)
\int_{\partial B} \frac{\partial v^{(1)}}{\partial \nu} \big|_{-} (\xi) \xi \, d\sigma(\xi), 
$$
with  $v^{(1)}$ being  such that
\begin{equation} \label{v1p2} \begin{cases}
\Delta_\xi v^{(1)} = 0 &  \mbox{in } \RR^d \smin \bar B,\\
 \Delta_\xi v^{(1)} = 0 &  \mbox{in }  B, \\
\dfrac{\pd v^{(1)}}{\pd \nu}|_+ = (\mu_m/\mu_c) \dfrac{\pd v^{(1)}}{\pd \nu} |_- &  \mbox{on } \pd B,\\
v^{(1)}(\xi) \sim \xi & \mbox{as } |\xi| \rightarrow +\infty.
\end{cases} \end{equation} 
The proof is then complete. \cqfd

\section{Shift of the scattering resonances by external small particles} \label{sec-3p2}

Now consider the case where the particle is outside $\O$. 
The main difference is that the modes of $K_\Omega^\omega$ are not defined on $D$, and therefore we must first write the expansion for $G$ outside of $\Omega$.
We start by recalling the Lippmann-Schwinger equation for $v=G-\Gamma_m$:
$$\left(I-\omega^2\tau \varepsilon_c \mu_m K_\Omega^\omega\right)[v(\cdot,x_0)](x)=\omega^2\tau \varepsilon_c \mu_m K_\Omega^\omega \left[\Gamma_m(\cdot,x_0) \right](x) \qquad \mbox{for } x,x_0\in \Omega .$$
Now, using Proposition \ref{prop:polepencilG} for $z$ and $z'$ inside $\Omega$ we have 
$$ v(z,z';\omega)=   c_{j_0}(\omega) \dfrac{e_{j_0}(z;\omega) e_{j_0}(z';\omega)}{\o-\o_0}+ \tilde{R}(z,z',\omega), $$ and we can write an expansion for $v(x,x_0)$ for $x\in \RR^d \setminus \Omega$:
\begin{multline*}
v(x,x_0) -\frac{ \omega^2\tau \varepsilon_c \mu_m c_{j_0}(\omega)}{\omega-\omega_0}\int_\Omega e_{j_0}(z,\omega)\Gamma_m(z,x)e_{j_0}(x_0,\omega) d z   - \omega^2\tau \varepsilon_c \mu_m K_\Omega^\omega[\tilde{R}(\cdot,x_0,\omega)](x)\\ = \omega^2\tau \varepsilon_c \mu_m K_\Omega^\omega \left[\Gamma_m(\cdot,x_0) \right](x) \qquad x \in \RR^d ,x_0\in \Omega.
\end{multline*}
The latter equality can be written as
\begin{align*}
v(x,x_0) = \frac{ \omega^2\tau \varepsilon_c \mu_m c_{j_0}(\omega)}{\omega-\omega_0}\left(\int_\Omega e_{j_0}(z,\omega)\Gamma_m(z,x) d z\right) e_{j_0}(x_0,\omega) + R_1(x,x_0,\omega), \qquad x \in \RR^d ,x_0\in \Omega.
\end{align*}
where $R_1$ is regular in space and holomorphic in $\omega$. 
Let  $$g_{j_0}(x;\omega) := \o^2 \tau \e_c \mu_m \int_\O e_{j_0} (z';\omega) \G_m(z,x;\omega) d z',\qquad x\in \RR^d .$$ We have
\begin{align*}
v(x,x_0) = \frac{ c_{j_0}(\omega)}{\omega-\omega_0}g_{j_0}(x;\omega) e_{j_0}(x_0,\omega) + R_1(x,x_0,\omega), \qquad x \in \RR^d ,x_0\in \Omega.
\end{align*}
We can now use this expansion in the Lippmann-Schwinger equation again:
\begin{multline*}
v(x,x_0) - \frac{ \omega^2\tau \varepsilon_c \mu_m c_{j_0}(\omega)}{\omega-\omega_0} g_{j_0}(x;\omega) \left(\int_\Omega e_{j_0}(z,\omega)\Gamma_m(z,x_0) d z\right)- \omega^2\tau \varepsilon_c \mu_m K_\Omega^\omega[R_1(\cdot,x_0,\omega)](x) \\ =  \omega^2\tau \varepsilon_c \mu_m K_\Omega^\omega \left[\Gamma_m(\cdot,x_0) \right](x) \qquad x \in \RR^d ,x_0\in \RR^d.
\end{multline*} Therefore, we have an expansion for $v$ outside of $\Omega$:
\begin{align*}
v(x,x_0) = \frac{ c_{j_0}(\omega)}{\omega-\omega_0}g_{j_0}(x;\omega) g_{j_0}(x_0;\omega) + R_2(x,x_0,\omega), \qquad x \in \RR^d ,x_0\in \RR^d.
\end{align*}
Analogously to the calculations in the previous section, we have
$$ v - \dfrac{c_{j_0} (\omega)}{\o - \o_0} ( v, \nabla g_{j_0})  L^{-1}[\nabla g_{j_0}] + L^{-1} R[v] = 0,$$
for some  operator $R$ with smooth kernel that is analytic in $\o$ in a neighborhood $V(\o_0)$ of $\omega_0$.
Therefore, by exactly the same method as in the previous section,  the following asymptotic expansion can be obtained. 
\begin{proposition} As $\delta \rightarrow 0$, we have
\begin{equation}
\o_\delta - \o_0 \simeq \delta^d c_{j_0}(\omega_0)  M(\mu_m  /\mu_c,B) \nabla g_{j_0}(z;\omega_0) \cdot \nabla {g}_{j_0} (z;\omega_0).  \end{equation}
\end{proposition}

\section{Shift of the scattering resonances due to resonant dispersive particles} \label{sec-4p2}

Let $D \Subset \O$ and suppose that $D$ is made of dispersive material, i.e., such that $\mu_c$ depends on $\omega$ and for a discrete set of frequencies $\omega$, that we can call plasmonic resonances by analogy with the transverse magnetic case,  problem (\ref{v1p2}) (or equivalently the operator $\displaystyle \big(\frac{\mu_m+\mu_c}{2(\mu_m - \mu_c)} I- K_D^* \big)^{-1}$)  is nearly singular, see \cite{plasmonic1,plasmonic2,plasmonic3}. In that case, we have the following scattering resonance  problem: Find $\omega$ such that there is a non-trivial solution $v$ to  
\begin{align}\label{eq:v}
L(\o) [v] - \dfrac{c_{j_0}(\omega)}{\o -\o_0} ( v, \nabla e_{j_0}) \nabla e_{j_0} + R[v] = 0 , 
\end{align}  
where $L(\o) = 1/( (\mu_m/\mu_c(\omega)) - 1) I - N_D^{0}$. Using the Drude model for the permeability, we
have $\mu_c(\o) = \mu_m (1 - \o_p^2 / \o^2)$, where $\o_p$ is the volume plasma frequency.

It is easy to see that the nearly singular character of (\ref{v1p2}) is linked to the 
non-invertibilty of $L(\omega)$ on ${W}$.

Denote by $P_1\ :\  L^2(D,\RR^d) \longrightarrow L^2(D,\RR^d) $ the orthogonal projector on $\nabla H^1_0(D)$ and $P_2\ : \  L^2(D,\RR^d) \longrightarrow L^2(D,\RR^d) $ the orthogonal projector on ${H}(\mathrm{div\ }0,D)$.
Using Lemma \ref{lem:orthdecomp}, we can write the resolvent operator $L^{-1}(\omega)$ as follows: $$L(\omega)^{-1} = \frac{1}{1-\lambda(\omega)} P_1 + \frac{1}{\lambda(\omega)}P_2+\sum_j \dfrac{( \cdot, \vphi_j) \vphi_j}{\l(\o) - \l_j},$$
where $(\lambda_j, \vphi_j)_j$ are the pairs of eigenvalues and associated orthonormal eigenfunctions of $N_D^0$. We can then rewrite equation \eqref{eq:v} as follows:
$$ v - \dfrac{c_{j_0}(\omega)}{\o - \o_0} \dfrac{( v , \nabla e_{j_0}) ( \nabla e_{j_0}, \vphi_j)  \vphi_j}{\l(\o) - \l_j} + L^{-1}R[v]= 0.$$
Now, taking the scalar product on $L^2(D,\RR^d)$ with $\nabla e_{j_0}$ and multiplying by $(\omega-\omega_0) (\lambda(\omega)-\lambda_j)$, we obtain that

\begin{align*}
(\omega-\omega_0) (\lambda(\omega)-\lambda_j) ( v,\nabla e_{j_0}) - c_{j_0} (\omega_0)  (v,\nabla e_{j_0})   (\nabla e_{j_0}, \vphi_j)^2 + (\omega-\omega_0) (\lambda(\omega)-\lambda_j) L^{-1}R[v] = 0.
\end{align*}
Since $R$ is analytic in $\omega$, the remainder $(\omega-\omega_0) (\lambda(\omega)-\lambda_j) L^{-1}R[v]$ is negligible in a neighborhood of $\omega_0$. Hence, we arrive at the following proposition:
\begin{proposition}
As $\delta \rightarrow 0$, we have
$$(\o_\delta - \o_0) (\l(\o_\delta) - \l_j) \simeq c_{j_0}(\omega_0) (\nabla e_{j_0}, \vphi_j)^2.$$
\end{proposition}

Note that if $\l(\o) - \l_j \simeq O(\o - \o_0)$ for $\o$ close to $\o_0$, then we obtain
$$(\o_\delta - \o_0)^2 \simeq c_{j_0}(\omega_0) ( \nabla e_{j_0}(\cdot; \omega_0), \vphi_j)^2,$$
Hence, we have a significant shift in the scattering resonances
if the particle $D$ is  resonant near or at the frequency $\o_0$. This anomalous effect has been observed in \cite{koenderink1}.

\section{Asymptotic analysis near exceptional scattering resonances} \label{sec-5p2}

In this section, we consider the asymptotic behavior of an exceptional scattering resonance for a particular form of the Green's function. These exceptional resonances are due to the non-Hermitian character of the operator $T_D^\omega$, see \cite{hai, excep1}. 
For simplicity and in view of the Jordan-type decomposition of the operator $T_D^\omega$ established in \cite{hai}, we assume that, for $\o$ near $\o_0$, $G(x,y;\omega)$ behaves like
\begin{equation}
\label{OpenTE:Gdecopm} G(x,y;\omega) = \G_m(x,y;\omega) + c_1(\omega) \dfrac{ h^{(1)}(x;\omega) h^{(1)}(y;\omega)}{\o - \o_0} + c_2(\omega) \dfrac{ h^{(2)} (x;\omega) h^{(2)}(y;\omega)}{(\o - \o_0)^2} + R(\o), \end{equation} 
 for two functions 
$h^{(1)}$ and $h^{(2)}$ in $\L^2(D)$. In this simple case, we characterize the shift of the scattering resonance $\omega_0$ due to the small particle $D$, which is assumed for simplicity to be non-plasmonic. 

Following the same arguments as those in the previous sections, we seek a non-trivial $v$ such that
$$L[v] - c_1(\omega) \dfrac{(v, \nabla h^{(1)})  }{\o - \o_0}\nabla {h^{(1)}} - c_2(\omega) \dfrac{(v, \nabla h^{(2)}) }{(\o - \o_0)^2} \nabla {h^{(2)}} = 0 ,$$
or equivalently, 
$$v - c_1(\omega) \dfrac{(v, \nabla h^{(1)}) }{\o - \o_0} L^{-1} [\nabla {h^{(1)}}]  - c_2(\omega) \dfrac{(v, \nabla h^{(2)})}{(\o - \o_0)^2} L^{-1}[\nabla {h^{(2)}}] = 0 .$$
By multiplying the above equation by $\nabla h^{(1)}$ and $\nabla h^{(2)}$, 
respectively, and integrating by parts over $D$, we obtain the following system of equations:
$$
\begin{cases}
(v, \nabla h ^{(1)}) \Bigg( 1 - c_1(\omega)  \dfrac{(L^{-1}[\nabla {h^{(1)}}] , \nabla h^{(1)}) }{\o - \o_0 } \Bigg)= c_2(\omega) (v, \nabla h^{(2)}) \dfrac{(L^{-1}[\nabla {h^{(2)}}], \nabla h ^{(1)})}{(\o - \o_0)^2},\\
(v, \nabla h^{(2)}) \Bigg(1 - c_2(\omega) \dfrac{(L^{-1}[\nabla {h^{(2)}}], \nabla h^{(2)})}{(\o - \o_0)^2} \Bigg) = c_1(\omega) (v, \nabla h ^{(1)})\dfrac{ (L^{-1}[\nabla {h^{(1)}}], \nabla h^{(2)})}{\o - \o_0}.
\end{cases}$$
Therefore, the following result holds.
\begin{proposition}
Assume that the decomposition (\ref{OpenTE:Gdecopm}) holds for $\omega$ near $\omega_0$. Then the perturbed scattering resonance problem (due to the particle $D$) can be reformulated as a search for $\o$ near $\o_0$ such that the matrix
$$\begin{pmatrix}
1 - c_1(\omega)  \dfrac{(L^{-1}[\nabla {h^{(1)}}] , \nabla h^{(1)}) }{\o - \o_0 } &
-c_2(\omega)\dfrac{(L^{-1}[\nabla {h^{(2)}}], \nabla h ^{(1)})}{(\o - \o_0)^2} \\
c_1(\omega) \dfrac{ (L^{-1}[\nabla {h^{(1)}}], \nabla h^{(2)})}{\o - \o_0} &
1 - c_2(\omega) \dfrac{(L^{-1}[\nabla {h^{(2)}}], \nabla h^{(2)})}{(\o - \o_0)^2} 
\end{pmatrix}$$
is singular.
\end{proposition}

\section{Concluding remarks} \label{sec-6p2}

In this paper, the leading-order term in the shifts of scattering resonances of a radiating dielectric cavity due to the presence of small particles is derived. The formula is in terms of the position and the polarization tensor of the particle. It is also proved that the shift is significantly enhanced if the particle is a plasmonic particle and resonates near or at a scattering resonance of the cavity. A characterization  of the shift due to small particles near an exceptional scattering resonance is performed. It would be challenging to develop a general theory near such frequencies. 

\appendix
\section{Proof of Proposition \ref{prop:polepencilG}} \label{appendix0}
\proof 
The proof follows an idea from \cite{hai}. Denote by $v$ the difference $$v(x,y)=G(x,y,\omega) - \Gamma_m(x,y,\omega).$$ One can check that $v(\cdot,x_0)$ is a solution of the following integral equation:
\begin{align*}
\left(I-\omega^2\tau \varepsilon_c \mu_m K_\Omega^\omega\right)[v]=\omega^2\tau \varepsilon_c \mu_m K_\Omega^\omega \left[\Gamma_m(\cdot,x_0) \right].
\end{align*}  Therefore, \begin{align*}
v= \left(\frac{1}{\omega^2\tau \varepsilon_c \mu_m} I - K_\Omega^\omega\right)^{-1}K_\Omega^\omega\left[\Gamma_m(\cdot,x_0)\right].
\end{align*}
Under the assumption that $\omega_0$ is a non exceptional scattering resonance (see Definition \ref{de:nonexceptionnal}) we can perform a pole pencil decomposition of the resolvent of $K_\Omega^\omega$. We start from the spectral decomposition of the compact operator $K_\Omega^\omega$ on $L^2(\Omega)$. The eigenspace associated with the eigenvalue $\frac{1}{\omega_0^2\tau \varepsilon_c \mu_m}$ is of dimension one, and we denote by $e_{j_0}$ its basis. One can then write
\begin{align*}
\left(\frac{1}{\omega^2\tau \varepsilon_c \mu_m}- K_\Omega^\omega\right)^{-1}=\frac{1}{\left(\omega^2\tau \varepsilon_c \mu_m\right)^{-1} - \lambda_{j_0}(\omega_0)} (e_{j_0}, \cdot )  e_{j_0} + \hat{R}(\cdot, \omega), 
\end{align*}
where $(\; ,\;)$ denotes the $L^2$ real scalar product on $\Omega$, and $\omega \mapsto \hat{R}(\cdot,\omega)\in L^2(\Omega)$ is analytic in a complex neighborhood $V$ of $\omega_0$. Using $$1 - \o^2 \tau \e_c \mu_m \l_{j_0} (\o) = R(\omega) (\o - \o_0)$$ and composing with $K_\Omega^\omega$, we obtain that 
\begin{align*}
v(x,x_0)= \tilde{c}_{j_0}(\omega)\frac{1}{\omega-\omega_0}  ({e}_{j_0},K_\Omega^\omega\left[ \Gamma_m(\cdot, x_0)\right]) e_{j_0} (x) + \tilde{R}(x,x_0,\omega). 
\end{align*}
Now we note that \begin{align*}
\Gamma_m(x,y) = - K_\Omega^\omega\left[\delta(\cdot - y)\right](x) \quad \mbox{for all } x,y \in \mathbb{R}^d, x\neq y. 
\end{align*}
Using the completeness relation given in Lemma \ref{lem:completeness} yields 
\begin{align*}
\Gamma_m(x,y) =\sum_j \lambda_j(\omega)e_j(y) e_j(x),
\end{align*}
for some constants$\lambda_j$. 
Now, we can write that
\begin{align*}
(e_{j_0},K_\Omega^\omega\left[ \Gamma_m(\cdot, x_0)\right])  = e_j(x_0) ( e_{j_0},\lambda_{j_0}(\omega)e_{j_0}), 
\end{align*}
to arrive at 
\begin{align*}
v(x,x_0)= c_{j_0}(\omega)\frac{1}{\omega-\omega_0}  e_{j_0}(x_0) e_{j_0} (x) + \tilde{R}(x,x_0,\omega).
\end{align*}
\cqfd

\section{Proof of Lemma \ref{lemap}} \label{appendix1}
\proof The operator $T_D$ is a singular integral operator of the Calderón-Zygmund type, see \cite{calderon}. This type of singular operator often arises in electrostatic and magnetostatic  theories (see the appendix of \cite{AmmariMillien2018} for a simple review of the properties of these operators within the formalism of Green's functions)
The fact that $T_D^\omega$ is well defined can be deduced directly from Proposition \ref{prop:polepencilG}. Since $G$ can be written as $G(x,y)= \Gamma_m(x,y) + K(x,y)$ where $K$ is a smooth kernel, we can see that the singularity of the derivatives of $G$ is the same as that of the derivatives of $\Gamma_m$, that is  $\partial_{x_i,x_j}G(x,y) = \partial_{x_i,x_j}\Gamma_m(x,y) + K_{i,j}(x,y)$. Therefore, it is easy to see that the singular part of $\partial_{x_i,x_j}G(x,y)$ satisfies the same cancellation property as $\partial_{x_i,x_j}\Gamma_m(x,y)$, that is, 
\begin{align*}
\int_{x+\mathbb{S}^{d-1}} \partial_{x_i,x_j}\Gamma_m(x,y)d y =0.
\end{align*} Hence, the fact that $T_D$ is defined on $L^2(D,\mathbb{R}^d)$ follows directly from classical Calderón-Zygmund theory and the cancellation property above. \cqfd
\section{Proof of estimate (\ref{eq:polarization})} \label{appendix2}

Here, we give some more details on how to obtain \eqref{eq:polarization} from the results of \cite{AmmariMillien2018}.
\begin{lemma} As $\delta \rightarrow 0$, we have
\begin{align*}
(L^{-1}[\nabla e_{j_0}], \nabla e_{j_0}) \simeq \delta^d M(\mu_m / \mu_c,B)  \nabla e_{j_0}(z; \omega_0) \cdot \nabla e_{j_0}(z; \omega_0).
\end{align*}
\end{lemma}
\proof 
From \cite[Proposition $3.1$]{AmmariMillien2018}, one can see that if $\varphi$ satisfies
\begin{align*}
\left\{\begin{aligned} \nabla \cdot \left(  \frac{1}{\mu} \nabla\varphi \right)= 0 \quad \mbox{in } \mathbb{R}^d, \\
\nabla \varphi(x) -e_{j_0}  = O\left( \vert x \vert^{-d+1}\right) \quad \mbox{as } |x| \rightarrow +\infty. \end{aligned}\right.
\end{align*} then $\nabla \varphi$ solves the integral equation
\begin{align*}
\left(\frac{1}{\mu_m} I - \left(\frac{1}{\mu_c}-\frac{1}{\mu_m}\right) N_D^0 \right)[\nabla \varphi]= \frac{1}{\mu_m} \nabla e_{j_0},  
\end{align*} which is exactly 
\begin{align*}
L[\varphi] = \nabla e_{j_0}.
\end{align*}
Now, replacing $\nabla e_{j_0}$ by its average and controlling the reminder via the Cauchy-Schwartz inequality we have:
\begin{align*}
( L^{-1}[\nabla e_{j_0}], \nabla e_{j_0})  =& ( L^{-1}[\nabla e_{j_0}], \frac{1}{|D|} \int_D \nabla e_{j_0}) + (L^{-1}[\nabla e_{j_0}], \nabla e_{j_0}- \frac{1}{|D|} \int_D \nabla e_{j_0}) \\
=& \frac{1}{|D|} \int_D L^{-1}[\nabla e_{j_0}] \cdot  \int_D \nabla e_{j_0}  + O\left(\delta^2\right).
\end{align*}
But the average of $\nabla \varphi$ is exactly the dipole moment, which is given by the polarization tensor applied to the average of the exciting field:
\begin{align*}
\int_D L^{-1}[\nabla e_{j_0}] =M(\mu_m / \mu_c,D)\int_D \nabla e_{j_0} = \delta^d M(\mu_m / \mu_c,B).
\end{align*}
Since $\frac{1}{|D|} \int_D \nabla e_{j_0}(x) dx - \nabla e_{j_0}(z) = O(\delta) $ (recall that $e_j$ is a mode of the cavity, and is therefore independant of $\delta$) we can replace the average of $\nabla e_{j_0}$ by its value at the center of $D$ to get the result. \cqfd

\end{document}